\documentclass[prl, twocolumn,showpacs,preprintnumbers,amsmath,amssymb]{revtex4}

\usepackage{graphicx}
\usepackage{bm}
\usepackage{amsmath, amssymb,amsfonts}

\usepackage{draftcopy} 
\begin{document}

\title{Non-invasive characterization of transverse beam emittance of electrons from a laser-plasma wakefield accelerator in the bubble regime using betatron x-ray radiation}

\author{S.~Kneip$^{1,2}$} 
\author{C.~McGuffey$^{2}$}
\author{J.~L.~Martins$^{3}$} 
\author{M.~Bloom$^{1}$}
\author{V.~Chvykov$^{2}$}
\author{F.~Dollar$^{2}$}
\author{R.~Fonseca$^{3,4}$}
\author{S.~Jolly$^{1}$}
\author{G.~Kalintchenko$^{2}$}
\author{K.~Krushelnick$^{2}$}
\author{A.~Maksimchuk$^{2}$}
\author{S.~P.~D.~Mangles$^{1}$}
\author{Z.~Najmudin$^{1}$}
\author{C.~A.~J.~Palmer$^{1}$}
\author{K.~Ta Phuoc$^{5}$}
\author{W.~Schumaker$^{2}$}
\author{L.~O.~Silva$^{3}$}
\author{J.~Vieira$^{3}$} 
\author{V.~Yanovsky$^{2}$}
\author{A.~G.~R.~Thomas$^{2}$}

\affiliation{$^1$The Blackett Laboratory, Imperial College London, London, SW7 2BZ, UK}
\affiliation{$^2$Center for Ultrafast Optical Science, University of Michigan, Ann Arbor, MI, 48109, USA}
\affiliation{$^3$ GoLP/Inst. Plasmas and Fus\~ao Nuclear, Laborat\'orio Associado, Instituto Superior T\'ecnico, Lisbon, 1049-001, Portugal}
\affiliation{$^4$ DCTI, ISCTE, Lisbon University Institute, Lisbon, 1649-026, Portugal}
\affiliation{$^5$Laboratoire d'Optique Appliqu\'ee, ENSTA, Ecole Polytechnique, Palaiseau, 91761, France}


\begin{abstract}
We propose and use a technique to measure the transverse emittance of a laser-wakefield accelerated beam of relativistic electrons. The technique is based on the simultaneous measurements of the electron beam divergence given by $v_{\perp}/v_{\parallel}$, the measured longitudinal spectrum $\gamma_\parallel$ and the transverse electron bunch size in the bubble $r_{\perp}$. The latter is obtained via the measurement of the source size of the x-rays emitted by the accelerating electron bunch in the bubble. We measure a \textit{normalised} RMS beam transverse emittance $<0.5$~$\pi$~mm$\:$mrad as an upper limit for a spatially gaussian, spectrally quasi-monoenergetic electron beam with 230~MeV energy in agreement with numerical modeling and analytic theory in the bubble regime.
\end{abstract} 
\maketitle
A laser wakefield accelerator \cite{TajimaT_PRL_1979} uses an intense laser pulse to generate a plasma wave, known as a wakefield, with a phase velocity close to the speed of light. If a relativistic electron beam can be inserted co-propagating with the wakefield, it can remain in an accelerating phase of the wave for a long distance, and extract substantial energy from the longitudinal electric field. Since plasma waves can support electric fields many orders of magnitude stronger than those in radio-frequency (RF) cavities, plasma accelerators provide a compact alternative to conventional RF linear accelerators in the future. 

In the nonlinear regime, the laser pulse expels electrons from its focal volume \cite{SunGZ_PFL_1987} but has a negligible effect on the remaining ions, resulting in a bare ion cavity with strong electromagnetic fields. Electrons from the periphery of the cavity are pulled inwards by the strong Coulomb attraction and form a thin high-density sheath around an approximately spherical ``bubble''  \cite{PukhovA_PPACF_2004,LuW_PRSTAAB_2007}. This has ideal linear accelerating and focusing properties for electrons within the cavity and at the point where the sheaths cross at the rear of the bubble, the field is particularly strong. Here, electrons from the plasma can be accelerated to the phase velocity of the bubble in a time shorter than their crossing time and are therefore trapped. Under certain conditions, the self-trapped electron beam can have a quasi-monoenergetic energy distribution \cite{ManglesSPD_Nature_2004,FaureJ_Nature_2004,GeddesCGR_Nature_2004}.

Most applications require the electron beam to be transported over long distances and/or to be focused into a small space with a minimum of divergence, for example to form high-resolution images. A good measure of the quality of a particle beam is its density in 6D phase space ($x,y,z,p_x,p_y,p_z$), which is a characteristic of the beam. The Liouville theorem states that for noninteracting particles, subject to a Hamiltonian (the electromagnetic fields of the plasma) the extent of the beam in phase space $\Gamma=\int p dr$, called emittance, is conserved. For practical purposes the emittance is split into 2D subspaces $x,p_x$, $y,p_y$ and $z,p_z$. If the area occupied by the beam in the $(r_{\perp},p_{\perp})$ plane is $\Gamma^\perp$ with $\perp=x,y$, then the \textit{normalised} or \textit{invariant} (transverse) emittance is defined
\begin{align}
\epsilon_n^{\perp}:=\Gamma^\perp/(\pi m_e c)
\label{equation1}
\end{align}
where $m_e$ is the electron mass and $c$ is the speed of light.
The gradients of the trajectory in the $(r_\perp,z)$ plane are more easily measured. This motivates the introduction of the trace space $(r_{\perp},r_{\perp}')$ where $r_{\perp}'=dr_{\perp}/dz=v_\perp/v_\parallel$. If the area occupied by the beam in trace space is $A^\perp$, then the geometric emittance is defined as
\begin{align}
\epsilon^\perp:=A^\perp/\pi
\label{equation2}
\end{align}
One can show that \textit{normalised} and \textit{geometric} emittance are related via $\epsilon_n^\perp=\epsilon^\perp \gamma \beta_z = \gamma \beta_z \int r_\perp'  dr_{\perp}\leq  \gamma \beta_z r_\perp'  r_{\perp}$ where $\beta_z=v_z/c$. The measurement of the \textit{normalised} emittance $\epsilon_n^\perp$ reduces to a measurement of the beam energy $\gamma$, the transverse extend $r_\perp$ and divergence of the beam $r_\perp'$, where $\beta_z\simeq1$ is assumed for relativistic beams.

The extent in $r_\perp$-space of a beam is relatively easy to obtain in general, e.g. by imaging its size with a scintillating screen. The extent in gradient $r_\perp'$-space of a  beam is more difficult to determine in general. For a lower energy electron beam, one standard measurement technique is the ``pepper-pot'' technique \cite{CollinsLE_NIM_1964,YamazakiY_1992_NIMPRSA}. A lead mask consisting of a series of small apertures, significantly smaller than the incident beam profile, allows the extent in $r_\perp$-space to be measured and to be related to the angular divergence $r_\perp'$ of the resulting beamlets at the aperture by assuming ballistic trajectories. This technique was applied to  $\simeq100$~MeV electron beams from a laser wakefield accelerator \cite{FritzlerS_PRL_2004,BrunettiE_PRL_2010,SearsCMS_PRSTAAB_2010} measuring \textit{normalised} transverse emittances of order $\pi$~mm~mrad. However it is difficult to extend this method to the GeV energy monoenergetic beams of recent laser wakefield experiments \cite{LeemansWP_NaturePhysics_2006,KneipS_PRL_2009} as, in addition to the difficulty of stopping GeV electrons, the spatial profile of the beam is small and therefore difficult to aperture. Another disadvantage of the ``pepper-pot'' technique is the difficulty of measuring the electron energy $\gamma$ simultaneously with $(r_\perp,r_\perp')$. The precision of the ``pepper-pot'' technique is ultimately limited by conflicting design considerations \cite{McDonaldKT_FrontiersParticleBeams_1989}.

In this Letter, we propose and use a technique that allows the simultaneous measurement of $(\gamma,r_\perp,r_\perp')$ and hence determination of an upper limit to the \textit{normalised} emittance $\epsilon_n^\perp$ on a single shot. The electron energy $\gamma$ is measured with a spectrometer, the divergence $r_\perp'$ is obtained from the distribution in the direction perpendicular to the dispersion and the beam size $r_\perp$ is inferred from the betatron x-ray source size emitted by the electron bunch. Using this technique, we show that the \textit{normalised} emittance of a $230$~MeV  electron beam from a LWFA can be $<0.5\,\pi \,\rm{mm}\,\rm{mrad}$ and agrees with the density scaling in the bubble regime. Our method will work without modification to measure the emittance of multi-GeV beams anticipated in the near future.

The experiments were performed at the \textsc{Hercules} laser of the Center for Ultrafast Optical Science at the University of Michigan, Ann Arbor. Horizontally, linearly polarised pulses with a central wavelength of $\lambda_L=800\,\rm{nm}$, a Gaussian full-width half maximum (FWHM) pulse duration of $t_L=(32\pm2)\,\rm{fs}$ and an energy of $\mathcal{E}_L=(2.2\pm0.1)\,\rm{J}$ were focused to a peak intensity of $I_0=(2.0\pm0.4)\times10^{19}\,\rm{Wcm}^{-2}$ or an $a_0=(3.0\pm0.3)$ with an off-axis parabolic mirror of focal length $f=1\,\rm{m}$ and an F-number of $F=10$. The transverse profile of the laser in vacuum yielded a FWHM focal diameter of $d_{FWHM}=(10.8\pm0.5)\,\mu \rm{m}$. Typically, $(27\pm2)\,\%$ of the laser energy is within the diameter $d_{FWHM}$. Helium gas was pulsed through a 3~mm diameter nozzle creating a supersonic jet. The density profile measured by interferometry has a FWHM length of $(2.7\pm0.2)\,\rm{mm}$ and consists of a central plateau, which is constant to within 10\%, and linear gradients at the edges. 

\begin{figure}
\begin{center}
\includegraphics[width=7.2cm]{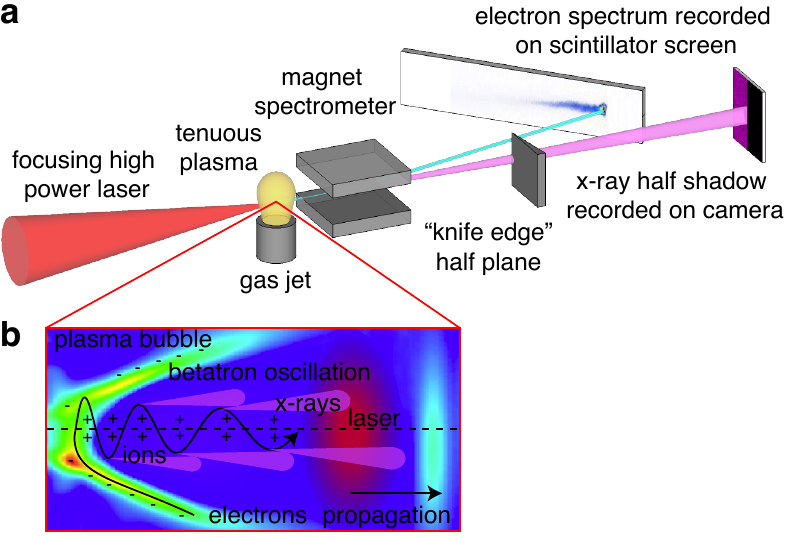}
\caption{Schematic of the (a) experimental setup and (b) inside of the plasma wakefield.}
\label{figure1}
\end{center}
\end{figure}

\begin{figure}[ht]
\begin{center}
\includegraphics[width=7.2cm]{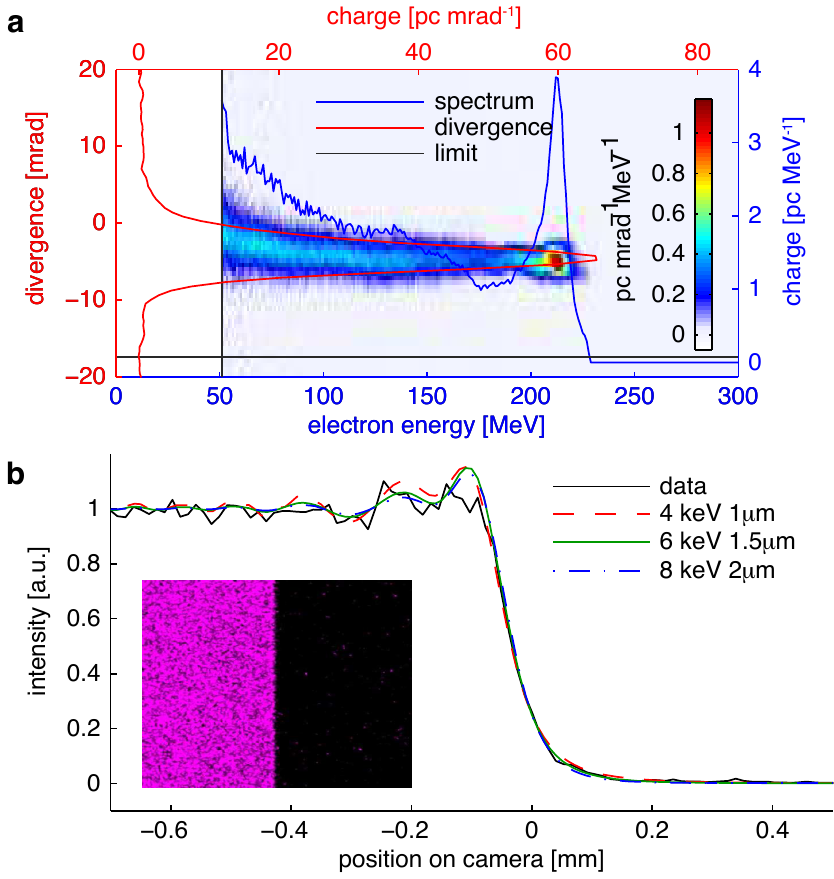}
\caption{(a) Sample electron data. Electron charge per mrad and MeV (2D image), and radially (red) and spectrally (blue) integrated curves. The dashed lines indicate the detector limit and noise level. (b) Sample x-ray half-shadow data (2D image), experimental x-ray intensity integrated along the edge (black line) and modeled x-ray intensity using Fresnel diffraction theory (dashed red, dotted green, dash-dotted blue lines).}
\label{figure2}
\end{center}
\end{figure}
\begin{figure}
\begin{center}
\includegraphics[width=7.2cm]{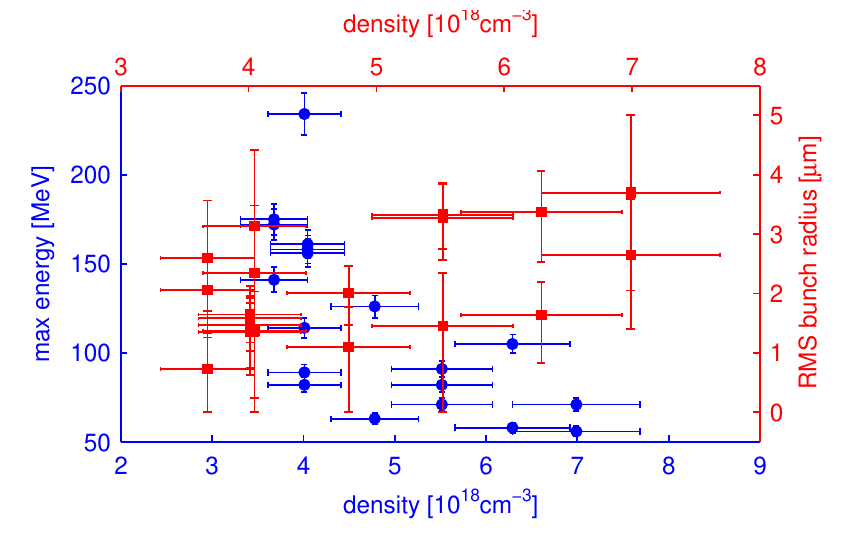}
\caption{Maximum electron energy (blue circles, left axis) and $1/e^2$ electron bunch radius (red squares, right axis) as a function of density. Horizontal error bars are $\pm10\%$ due to uncertainty in density determination and vertical error bars are due to uncertainty in electron energy ($\pm 5\%$) and x-ray source size determination ($95\%$ confidence interval).}
\label{figure3}
\end{center}
\end{figure}

Figure \ref{figure1}a shows a schematic of the experimental setup. A permanent magnet deflects the electrons from the laser axis and disperses them horizontally according to their energy onto a scintillator screen, which is imaged with a CCD camera to obtain the electron spectrum and electron beam divergence. Electrons with energies between 50 and 500~MeV and divergence angles between $\pm10\,\rm{mrad}$ can be detected. Electron beams are observed for electron densities between $3$ and $8\times10^{18}\,\rm{cm}^{-3}$. A sample electron beam is shown in figure \ref{figure2}a. The average beam charge is found to be $130\pm100\,\rm{pC}$. The maximum achievable electron beam energy was found to increase from 70 to 230~MeV for densities from $8$ to $3\times10^{18}\,\rm{cm}^{-3}$. Figure \ref{figure3} shows the scaling of the electron energy with density. At low densities, electron beams were characterised by a high energy mono-energetic features, with a low energy dark current (fig.~\ref{figure2}a). For increasing densities, electron beams developed a more complex poly-energetic structure and eventually a 100\% energy spread.

The propagation of the electrons is not affected by the magnet in the direction perpendicular to the dispersion of the spectrometer and thus the vertical electron beam divergence can be obtained from the 1D dispersed signal (fig.~\ref{figure2}a). Since beams are not necessarily mono-energetic, we integrate the dispersed spectrum along the energy axis to obtain an average RMS (root mean square) beam divergence.
We find that the divergence shows little correlation with the density but may increase somewhat with higher densities. This was was observed before \cite{ManglesSPD_PRL_2006,KneipS_PPCF_2011}.

It has been shown that electrons accelerated in a plasma wave produce copious amounts of synchrotron-like radiation, termed betatron x-rays \cite{RousseA_PRL_2004,KneipS_PRL_2008}. This is due to the electrons performing transverse oscillations whilst being accelerated along the laser axis in the fields of the plasma bubble, as illustrated in figure \ref{figure1}b. 
The characteristics of the betatron x-rays are related to the properties of the electron beam and plasma wiggler. In particular the small x-ray source size can serve as a diagnostic for the transverse electron bunch size. With our choice of laser parameters, the spot size is matched to the bubble radius $w_0=r_b=2\sqrt{a}\cdot c / \omega_p$ for densities of $3.3-4.9\times10^{18}\,\text{cm}^{-3}$ and the pulse duration is matched to the bubble radius $c \cdot t_L=r_b$ for densities of $2.9-4.6\times10^{18}\,\text{cm}^{-3}$. If these conditions are fulfilled, the wakefield accelerator operates in the bubble regime \cite{LuW_PRSTAAB_2007}. Dephasing then coincides with depletion and maximises the electron beam energy and thus the x-ray power at the end of the interaction. The source size of the x-rays therefore gives the transverse bunch size $r_\perp$ of the electron beam at the end of the interaction.

As shown in figure \ref{figure1}a, penumbral imaging is used to determine the betatron x-ray source size. A half-plane (thinly cleaved InSb crystal) is placed after the x-ray source, casting a half-shadow on the detector. The experimental intensity distribution integrated along the edge is shown in figure \ref{figure2}b and can be modeled with Fresnel diffraction theory, yielding x-ray spectrum and source size \cite{KneipS_NaturePhysics_2010}. Using a least-squares method, the RMS x-ray source radius (transverse electron bunch radius) is calculated with $95\%$ confidence interval, as shown in figure \ref{figure3}, right axis. Much like the divergence, the x-ray source size also shows little correlation with density.

\begin{figure}
\begin{center}
\includegraphics[width=7.2cm]{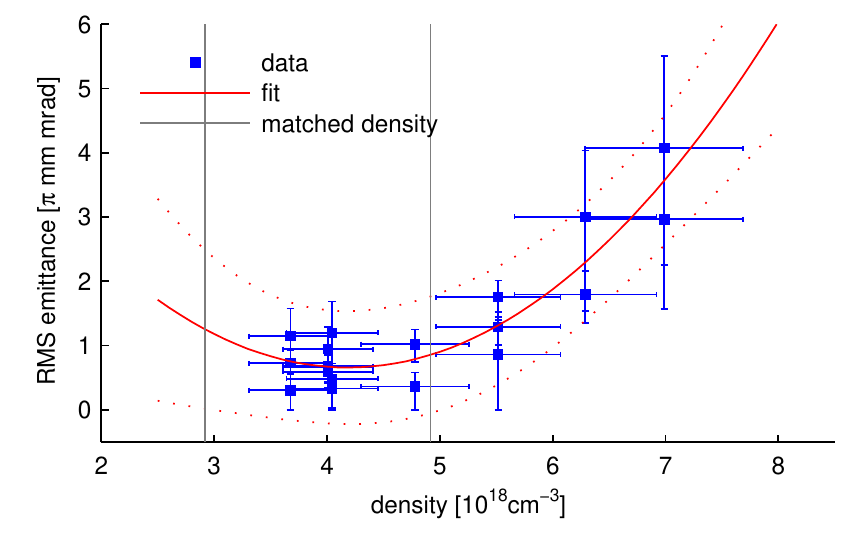}
\caption{Experimentally determined emittance as function of density (blue squares). Fit with a polynomial of 2nd order (red solid line), giving $R^2=0.85$ and $95\%$ confidence interval (dashed red). The vertical grey lines indicate the density interval for which the spot size $w_0=r_b$ and pulse duration $c \cdot t_L=r_b$ matches the bubble radius.}
\label{figure4}
\end{center}
\end{figure}

Using equ.~\ref{equation1} and \ref{equation2}, the measured $\gamma_{||}$, $v_{\perp}/v_{||}$ and $r_{\perp}$ can be combined to obtain an upper limit for the RMS transverse normalized emittance, which is plotted in figure \ref{figure4}. Despite the fact that each measured quantity shows little correlation with density, the emittance gives a correlation of $C=0.84$ with density. The associated p-value (probability of getting a correlation as large as the observed value by random chance, when the true correlation is zero) is $P<10^{-4}$. A quadratic fit gives an $R^2=0.83$.

Two effects will determine the scaling of the emittance. Firstly, for constant laser pulse duration, the emittance will grow at high densities, as the electron beam is more likely to interact with the laser pulse \cite{ManglesSPD_PRL_2006,KneipS_PRL_2008}. This should become significant as $c \cdot t_L> r_b$, i.e. for densities $>2.9-4.6\times10^{18}\,\text{cm}^{-3}$. This trend is reflected in our data (figure \ref{figure4}). Secondly, the emittance will scale with the bubble size $\propto n_e^{-1/2}$ and the transverse electron momentum spread $\propto a_0$ \cite{LuW_PRSTAAB_2007}, leading to an increase at low densities. This trend is not reflected well by our data, which does not extend to densities much lower than where $c \cdot t_L\simeq r_b$ (shaded area in figure \ref{figure4}). The fit to our data however suggests that the emittance is minimized at $n_e\simeq 4 \times 10^{18}\,\text{cm}^{-3}$, which is when $w_0=r_b$ and $c \cdot t_L=r_b$ are matched. 

An analytic derivation for the bubble regime using the FWHM quantities for $(r_{\perp},p_{\perp})$ gives the FWHM transverse \textit{normalized} emittance as $\epsilon_{n}^\perp=0.033\,a_0^{3/2}(n_{\text{crit}}/n_e)^{1/4}$, which computes to $\epsilon_n^\perp \simeq0.87$ and 
$1.59\,\pi\,\rm{mm}\,\rm{mrad}$ for $n_e=4\times10^{18}\,\rm{cm}^{-3}$ and 
$a=3.2$ and $4.8$ respectively \cite{ThomasAGR_PoP_2010}. The latter value is the self-consistent trapping threshold for the given density. The predicted emittance is in very good agreement with the measured RMS normalized transverse emittance $\simeq0.5\,\pi\,\rm{mm}\,\rm{mrad}$ at $n_e=4\times10^{18}\,\rm{cm}^{-3}$.

In computing the measured emittance, we used the spectrally averaged divergence $v_{\perp}/v_{\parallel}$ and radially averaged mean energy $\gamma_{\parallel}$. The radiated x-ray power per solid angle is $\propto \gamma_{\perp}^4$ which affects the measured x-ray source size and electron bunch size $r_{\perp}$. This may motivate a weighing of the 2D electron data with $\propto \gamma_{\perp}^4$ before computing averaged $v_{\perp}/v_{\parallel}$ and  $\gamma_{\parallel}$. When done, we find that both the absolute number and density trend of the emittance differs insignificantly from our figure \ref{figure4}.

Comprehensive numerical modelling was performed to verify if such small emittances can occur in the bubble regime. Three-dimensional particle in cell simulations with the code OSIRIS \cite{FonsecaRA_LNICS_2002} were carried out, in which a pulse with experimental parameters ($a_0=3.0$, $t_L=32$~fs, $d_{\text{FWHM}}=10.8\,\mu$m) is focused 0.25~mm into the plasma. The electron plasma density profile increases linearly from zero to $n_e = 6\times 10^{18}~$cm$^{-3}$ in the first 0.5~mm, it is constant for 2.7~mm, and falls linearly to zero in 0.5~mm. To save computational time, simulations were performed in a relativistically boosted frame ($\gamma = 5$) \cite{MartinsSF_NaturePhysics_2010}. The simulation box corresponds to $3.5\times0.11\times0.11$~mm$^3$ in the laboratory frame. A total of $9\times10^9$ particles were pushed for $\approx6\times 10^3$ iterations. The resolution in the laser propagation direction (in the boosted frame) is $k_0\Delta z = 0.12$ and $k_p\Delta r_{\perp} = 0.16$ in the transverse directions.
\begin{figure}
\begin{center}
\includegraphics[width=7cm]{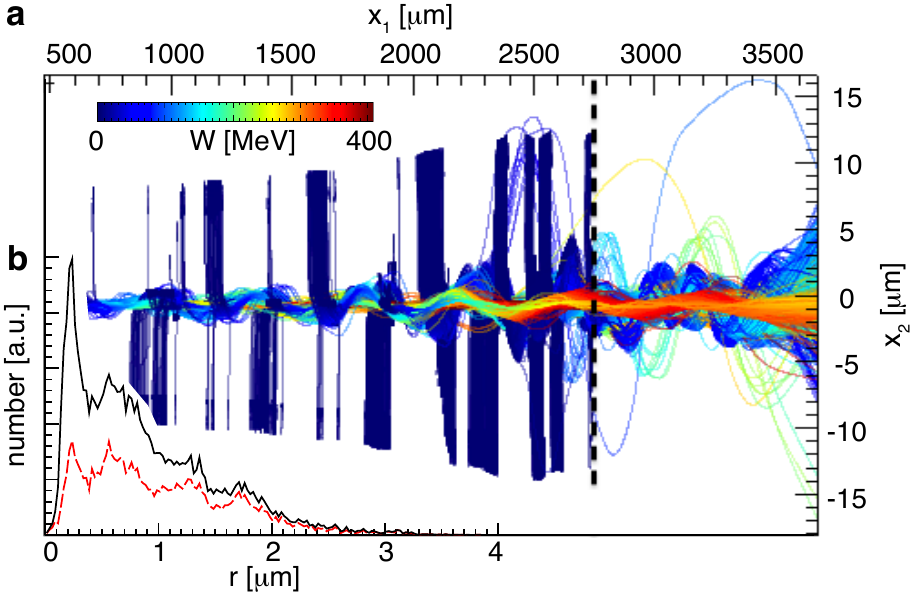}
\caption{(a) Electron trajectories obtained from the modeling. (b) Histogram of the betatron amplitudes for entire 3.7~mm jet (solid black) and for the last 1~mm (dashed red).}
\label{figure5}
\end{center}
\end{figure}
The modeling predicts a polyenergetic electron beam with peaks at 210 and 300~MeV, an average energy of 240~MeV and an almost symmetric profile with divergence $v_{\perp}/v_{\parallel}\simeq4.6$~mrad. As the code tracks the particle trajectories we find that electrons are injected throughout the interaction and all of them perform oscillating trajectories as depicted in figure \ref{figure5}a. We analyze the oscillation radius for each electron and oscillation to obtain a histogram of oscillation radii, shown in figure \ref{figure5}b. From the histogram we obtain an average oscillation amplitude or bunch radius of $r_{\perp}=0.9$ and $1.0\,\mu \rm{m}$ respectively, depending on whether we average over the entire length of the trajectory or the last 1~mm. This computes to a transverse normalized emittance of $\epsilon_{\perp}^n=0.6-0.7\,\pi\,\rm{mm}\,\rm{mrad}$ which is in good agreement with the measurement and analytical calculation.

We have proposed and used a technique to measure the emittance of a LWFA electron beam. The method infers the transverse electron bunch size in the bubble from the measurement of the betatron x-ray source size. With this method we find that the transverse \textit{normalised} emittance of a 230~MeV electron beam in the bubble regime can be $\epsilon_n^i<0.5\,\pi\,\rm{mm}\,\rm{mrad}$ similar to conventional accelerators. The method can be applied to multi-GeV electron beams anticipated from LWFA in the near future, where standard single shot emittance measurements have limited accuracy and eventually fail. Our method will be crucial to better diagnose LWFA and help to progress them towards applications.

This work was supported by U.S. NRC grant no. 38-09-953. The work of J.L.M., R.F., J.V. and L.O.S. is partially supported by FCT (Portugal). The simulations were performed on the IST Cluster (Portugal) and on Jugene (Germany).

\end{document}